\documentstyle[twocolumn,psfig]{jpsj}

\def\runtitle{Orbital symmetry of a triplet pairing  UPt$_3$
} 
\def\runauthor{Kazushige {\sc Machida} Tsutomu 
{\sc Nishira} and Tetsuo {\sc Ohmi}}

\title
{
Orbital symmetry of a triplet pairing 
in a heavy Fermion superconductor UPt$_3$
} 

\author
{  
Kazushige {\sc Machida},$^{1}$\footnote{e-mail: 
machida@mp.okayama-u.ac.jp} Tsutomu {\sc Nishira}$^{1}$ and Tetsuo {\sc Ohmi}$^{2}$\footnote{e-mail: ohmi@ton.scphys.kyoto-u.ac.jp}
}

\inst
{$^1$Department of Physics, Okayama University,
         Okayama 700-8530, \\  \ 
$^2$Department of Physics, Graduate School of Science, Kyoto University, Kyoto 
606-8502}

\abst
{
The orbital symmetry of the superconducting 
order parameter in UPt$_3$
is identified by evaluating the directionally dependent 
thermalconductivity and ultrasound attenuation in the 
clean limit
and compared with the existing data for both basal plane and the
$c$-axis of a hexagonal crystal.
The resulting two component orbital part expressed by 
$(\lambda_x({\bf k}), \lambda_y({\bf k}))$ is combined with the 
previously determined triplet spin part, leading to 
the order parameter of either the non-unitary bipolar state of the form: ${\bf d}({\bf k})={\hat {\bf b}}\lambda_x({\bf k})+i{\hat {\bf j}}\lambda_y({\bf k})$ or the unitary planar state of the form: ${\bf d}({\bf k})={\hat {\bf b}}\lambda_x({\bf k})+{\hat {\bf j}}\lambda_y({\bf k})$ where
${\hat {\bf b}}\perp{\hat {\bf j}}={\hat {\bf c}}$, or ${\hat {\bf a}}$ with the 
hexagonal unit vectors ${\hat {\bf a}}, {\hat {\bf b}}$
and ${\hat {\bf c}}$. The $d$ vector 
is rotatable in the plane spanned by ${\hat {\bf a}}$ and ${\hat {\bf c}}$ perpendicular to ${\hat {\bf b}}$  under weak applied $c$-axis 
field because of the weak spin
orbit coupling. Experiments are proposed to distinguish between 
the equally possible these states.
}

\kword
{
heavy Fermion superconductor, UPt$_3$, triplet pairing, orbital symmetry, thermalconductivity, ultrasound attenuation, bipolar state,
planar state
}

\begin{document}
\sloppy

\maketitle

\section{Introduction}

Much attention has been focused on various unconventional superconductors in the recent years, including high $T_c$  cuprates, ruthenate, low dimensional organic conductors and heavy Fermion materials. Extensive experimental and theoretical studies on UPt$_3$, 
a typical heavy Fermion superconductor, have been devoted to in order to elucidate the pairing symmetry realized in this particular material\cite{heffner}.
UPt$_3$ discovered in 1983\cite{stewart} is rather special in this class of materials because it is found in 1988\cite{fisher} that the superconducting transition temperature is split ($T_{c1}\sim$0.58K and $T_{c2}\sim$0.53K), self-evidently suggesting a rich internal structure of the Cooper pair, definitely not a simple
conventional s-wave pairing. There are at least three phases identified: 
A (high $T$ and low $H$), B (low $T$ and low $H$) and C (high $T$ and high $H$) phases. In spite of these facts and the following much efforts, 
however, it was a difficult task to identify the exact pairing symmetry.

Vast majority of experimental works\cite{heffner} done on UPt$_3$ are
concerned with the energy gap structure in the superconducting state 
by measuring the $T$-dependences of thermodynamic and transport quantities whose power law behaviors at low $T$'s point to the existence of at least a line node at the equator of the Fermi surface. Precise gap structure in addition to
that line node, however, remains still to be worked out.
At this stage of the research there were three major scenarios by 
Joynt
\cite{joynt}, Sauls\cite{sauls1,sauls2,sauls3} and ours\cite{our1,our2,our3} who try to consistently explain the above $T_c$-splitting phenomenon,
other than the so-called accidental degeneracy model.\cite{garg}

In general, the pairing symmetry consists of the orbital part
and the spin part in the pairing function: ${\hat\Delta}_{\alpha,\beta}({\bf k})=i({\vec \sigma}\cdot{\bf 
d}({\bf k})\sigma_2)_{\alpha,\beta}$. The former determines the gap topology which the above measurements probe. The latter characterizes the spin structure of the Cooper pair, which is related to the spin susceptibility of the system, is usually probed by the Knight shift experiment of NMR.

A long waited Knight shift experiment by using high quality single crystals has been finally performed by Tou et al\cite{tou,nmr}, revealing the rich details of the spin structure of the pairing function in UPt$_3$:
(1) The spin susceptibility $\chi_i$  for the $i$-direction  
decreases below $T\le T_{c2}$ for $i=b$ and $c$-axes of the hexagonal
crystal at low fields, corresponding to the B phase.
(2) The parity of the pairing function is odd and the spin triplet state is realized. 
(3) The low temperature and low field state B phase is characterized by having the two-component  $d$-vector.
By increasing $H$ ($\parallel c$) one of the $d$-vector component rotates to lower the Zeeman energy around 2kG,  implying that the effective spin-orbit coupling 
which is to pin the $d$-vector to the crystal lattice is weak. These experimental facts\cite{tou,nmr} fit to our predicted scenario\cite{our1,our2,our3}, negating the E$_{2u}$
scenario due to Sauls\cite{sauls1,sauls2} who assumes the one-component $d$-vector  fixed to the $c$-axis by strong spin-orbit coupling and also ruling out the E$_{1g}$
scenario due to Joynt\cite{joynt} who predicts the spin susceptibility to decrease for all $H$-orientations. This is not observed by Tou et al's NMR experiment\cite{nmr}.

So far we have not so seriously considered the orbital structure of the pairing function partly because without identifying its spin structure it was difficult to narrow down the precise orbital symmetry among vast possible candidates. Now since its spin structure, we believe, has fairly well determined, we must select out the orbital part to complete our task.

For this purpose we calculate the $T$-dependences of the thermalconductivity and ultrasound attenuation coefficient, each for major crystalline orientations of the $c$-axis and the basal plane for the two candidate orbital states whose spin structures are identical, namely, the non-unitary bipolar state and the unitary planar state. By adjusting several material parameters we try to fit the data of the above quantities in a consistent manner.

Arrangement of the paper is as follows: We first introduce the two candidate orbital states, both of which can equally explain the above Knight shift experiment  and thus logically same in this respect. In next Section we calculate the thermalconductivity and ultrasound attenuation coefficient by using these states. In the Sec.4 
we explain the phase transitions in terms of these states. The final Section is devoted to discussions and conclusions which summarize our present status for the identification of the pairing symmetry done over the past ten years\cite{our1,our2,our3}.

\section{Non-unitary bipolar state and unitary planar state}
\subsection{Choice of the order parameters}

The inert $p$-wave states with odd-parity under the hexagonal symmetry
are enumerated group-theoretically when the spin-orbit coupling is 
weak\cite{ozaki}, which is the present case.
An extension to more higher order angular momentum state
is trivial, simply replacing the irreducible two-dimensional representation whose basis function is given by $(k_x,k_y)$ for the $p$-wave case in the cubic symmetry to the 
corresponding two-dimensional E$_{1u}$ and E$_{2u}$ representations in the hexagonal symmetry
characterized by $(\lambda_x({\bf k}), \lambda_y({\bf k}))$.
The enumerated six states  are listed in Table 1.

\begin{table}
\caption{The enumerated states belonging to the 
two-dimensional representations characterized by 
$(\lambda_x({\bf k}), \lambda_y({\bf k}))$ where ${\vec \tau}=i{\vec \sigma}\sigma_2$.}
%
% \begin{tabular}{ccc}
% &state& $d$ vector\\
% Unitary&&\\
%   &Axial state&$\tau_z(\lambda_x+i\lambda_y)$\\
%   &Planar state&$\lambda_x\tau_x+\lambda_y\tau_y$\\
%   &Polar state&$\lambda_x\tau_z$ or $\lambda_y\tau_z$\\
% Non-unitary&&\\
%   &$\beta$ state&$(\tau_x+i\tau_y)\lambda_x$, or $(\tau_x+i\tau_y)\lambda_y$\\
%   &Bipolar state&$\lambda_x\tau_x+i\lambda_y\tau_y$\\
%   &$\gamma$ state&$(\tau_x+i\tau_y)(\lambda_x+i\lambda_y)$\\
% \end{tabular}
\begin{tabular}{@{\hspace{\tabcolsep}\extracolsep{\fill}}ccc} \hline \hline
&state& $d$ vector\\       \hline
Unitary&&\\
  &Axial state&$\tau_z(\lambda_x+i\lambda_y)$\\
  &Planar state&$\lambda_x\tau_x+\lambda_y\tau_y$\\
  &Polar state&$\lambda_x\tau_z$ or $\lambda_y\tau_z$\\
Non-unitary&&\\
  &$\beta$ state&$(\tau_x+i\tau_y)\lambda_x$, or $(\tau_x+i\tau_y)\lambda_y$\\
  &Bipolar state&$\lambda_x\tau_x+i\lambda_y\tau_y$\\
  &$\gamma$ state&$(\tau_x+i\tau_y)(\lambda_x+i\lambda_y)$\\ \hline \hline
\end{tabular}
\label{table1}
\end{table}

Among these states listed in Table 1 several requirements in light of the UPt$_3$
problem should be met to select. The decisive one is that the spin component must be two according to the above Knight shift experiments. 
Previously we advocated a $\beta$ type state or a 
$\gamma$ type state\cite{our1,our2,our3}, 
but these turn out not to be adequate by the following two reasons:
(A) The $\mu$SR experiment by Luke et al\cite{luke} claims that the spontaneous moment appears below $T_{c2}$, suggesting the time reversal symmetry breaking state realized as the B phase. 
The non-unitary $\beta$ state  is consistent with this, but not the planar and bipolar.
However, no later experiments reproduce their result\cite{others}.
(B) The following analysis of the thermodynamic and transport data also shows that the $\beta$ 
and $\gamma$ states are not appropriate in explaining their $T$-dependences.
Then we are left with the only two states: the bipolar and planar states in Table 1.
The former is non-unitary while the latter is unitary. Both 
have no spontaneous moment averaged over the Fermi surface.
We proceed in the following sections with these states to calculate physical quantities.
In Appendix we provide a proof that these particular states are local minimum and thus stable solutions of the general Ginzburg-Landau type free energy functional 
under a certain parameter space spanned by $\beta_1\sim\beta_5$
because it is known that the bipolar and planar states with three-component orbital part are saddle point solutions\cite{jones}. Then we assume that either bipolar state
or planar state is stabilized by fulfilling the above conditions (see Appendix).

\subsection{Gap function and gap topology}

The spin triplet pairing function is described by a $2\times 2$ form
\begin{eqnarray}
\hat{\Delta}_{\alpha\beta}({\bf k})= \left(
    \begin{array}{cc}
           -d_x({\bf k})+id_y({\bf k}) & d_z({\bf k}) \\
            d_z({\bf k}) & d_x({\bf k})+id_y({\bf k}) 
    \end{array}
  \right) \label{eq:Delta}
\end{eqnarray}
in terms of the ${\bf d}({\bf k})=(d_x({\bf k}),d_y({\bf k}),d_z({\bf k}))$ vector
which fully characterizes each state ($d_i({\bf k})$ is a complex number).
The gap function is given by
\begin{eqnarray}
\Delta ({\bf k})\propto \sqrt{|{\bf d}({\bf k})|^2\pm|{\bf d}({\bf k})\times{\bf d}^{\ast}({\bf k})|}.\label{eq:gap1}
\end{eqnarray}
If ${\bf d}({\bf k})$ is non-unitary, the gap function has two branches, each corresponding to the spin up and down pairs.
If ${\bf d}({\bf k})$ is unitary, the gap function has only 
one degenerate branch given by 
\begin{eqnarray}
\Delta({\bf k})\propto |{\bf d}({\bf k})|.\label{eq:gap2}
\end{eqnarray}
We call the bipolar state belonging to a non-unitary state\cite{simplenon-u} when 
${\bf d}({\bf k})=(\lambda_x({\bf k}), i\lambda_y({\bf k}), 0)$ and the
planar state belonging to a unitary state when
${\bf d}({\bf k})=(\lambda_x({\bf k}), \lambda_y({\bf k}), 0)$
where the two kinds of the orbital functions 
$\lambda_x({\bf k})$,  and $\lambda_y({\bf k})$ are real number functions.
The orbital function $(\lambda_x({\bf k}), i\lambda_y({\bf k}))$ must
belong to two-dimensional irreducible representations in D$_{6h}$,
namely, E$_{1u}$\cite{E1u}, or E$_{2u}$. Since there must be
a line node on the equator of the Fermi surface,
we are bound to choose the orbital function in E$_{2u}$ given by
\begin{eqnarray}
\lambda_x({\bf k}) &=& k_z(k_x^2-k_y^2)G_{total}({\bf k})    \\       \label{eq:lambdax}
\lambda_y({\bf k}) &=& k_z2k_xk_yG_{total}({\bf k})       \\       \label{eq:lambday}
G_{total}({\bf k}) &=& (1+a_2k_z^2+a_4k_z^4)        \nonumber \\
&& 
\{1+b_2(k_x^2+k_y^2)^2+b_4(k_x^2+k_y^2)^4\}.                \label{eq:total}
\end{eqnarray}
These are supplemented with a totally symmetric function
 $G_{total}({\bf k})$ parameterized by $a_2$, $a_4$, $b_2$, and $b_4$.

The actual gap topology for the bipolar state is easily found from Eq. (2);
in the polar coordinates $(\theta,\phi)$ on the Fermi sphere
\begin{eqnarray}
&& \!\!\!
\sqrt{\lambda_x^2({\bf k})+\lambda_y^2({\bf k})\pm 2\lambda_x({\bf k})\lambda_y({\bf k})}  \nonumber 
\\ &&
 \qquad \qquad = |\cos\theta|\sin^2\theta\sqrt{1\pm\sin 4\phi},          \label{eq:topology1}
\end{eqnarray}
indicating that (1) four line nodes parallel to the $k_z$-axis, (2) one
line node on the equator  in the absence of $a_2$, $a_4$, $b_2$, and $b_4$,
which are chosen to tune the fine details  of the nodal topology.
(3) Since in the A and C phases one of the two components vanishes,
there are one line node on the equator and two line nodes parallel to
the $k_z$-axis whose positions are different for the A and C phases.

In the unitary planar state the gap topology\cite{same} is determined by
\begin{eqnarray}
\sqrt{\lambda_x^2({\bf k})+\lambda_y^2({\bf k})}=|\cos\theta|\sin^2 \theta,\label{eq:topology2}
\end{eqnarray}
in the absence of $a_2$, $a_4$, $b_2$, and $b_4$.
This shows (1) the two quadratic point nodes at the poles ($\theta=0,\pi$), 
(2) a linear line node on the equator ($\theta=\pi/2$),
and (3) in the A and C phases there are either two point nodes or one line node.

The $T$-dependence of the gap function $\Delta_{\pm}^2({\bf k}, T)$
for the bipolar state is given by 
\begin{eqnarray}
\Delta_{\pm}^2({\bf k}, T)&=&\Delta_x^2({\bf k}, T)+\Delta_y^2({\bf k}, T) \nonumber \\
&&\pm 2\Delta_x({\bf k}, T)\Delta_y({\bf k}, T)            \label{eq:Tdep1}
\end{eqnarray}
where
\begin{eqnarray}
\Delta_x({\bf k}, T) &=& \Delta^{(x)}_0({T\over T_{c1}})\lambda_x({\bf k})\label{eq:Tdep2}   \\
\Delta_y({\bf k}, T) &=& \Delta^{(y)}_0({T\over T_{c2}})\lambda_y({\bf k}).\label{eq:Tdep3}
\end{eqnarray}
The order parameter amplitudes $\Delta^{(x)}_0(T)$ and $\Delta^{(y)}_0(T)$, where 
due to the T$_c$ splitting ($T_{c2}/T_{c1}=0.9$) each amplitude appears either
at $T=T_{c1}$ or $T=T_{c2}$, are evaluated by solving the BCS type gap equations.

Similarly, in the planar state case the gap function is 
\begin{eqnarray}
\Delta^2({\bf k}, T)=\Delta_x^2({\bf k}, T)+\Delta_y^2({\bf k}, T)\label{eq:Tdep4}
\end{eqnarray}
instead of Eq.\ (\ref{eq:Tdep1}) in the bipolar state.
The order parameter amplitudes $\Delta^{(x)}_0(T)$ and $\Delta^{(y)}_0(T)$ are
given by Eqs.\ (\ref{eq:Tdep2}) and \ (\ref{eq:Tdep3}).

\section{Thermalconductivity and ultrasound attenuation}

In order to identify the proposed two states, and to see which is better,
we calculate their transport properties:
the thermalconductivity and ultrasound attenuation coefficient taking into account
resonance scattering effects in the unitarity limit where the phase shift $\delta_0=\pi/2$.
The formulation for these quantities  just follows those by Hirschfeld
et al\cite{hirschfeld} and by Norman and Hirschfeld\cite{norman} who extend
the former to considering inelastic scattering effects important in high temperatures
near $T_c$. We are interested in the clean limit behavior here.

The thermalconductivity $\kappa_i(T)$ for the heat current along the
$i$-direction is given by
\begin{eqnarray}
{\kappa_i(T)\over \kappa_N(T_{c1})} &=& {9\over 2\pi^2T_c}
\int^\infty_0 d\omega ({\omega\over T})^2 {\rm sech}^2{\omega\over 2T}K_i(\omega, T)\label{eq:kappa}
\\
K_i(\omega,T) &=&
{\langle{\hat k_i^2{\rm Re}\sqrt{\omega^2-|\Delta({\bf k},T)|^2}}\rangle
\over
{\langle{\rm Re}{1\over \sqrt{\omega^2-|\Delta({\bf k},T)|^2}}\rangle}}
\label{eq:k}
\end{eqnarray}
where $\kappa_N(T_{c1})$ is the thermalconductivity of the normal 
state at $T=T_{c1}$. The average over the Fermi surface is denoted as $\langle\cdots\rangle$. 
We assume the Fermi sphere in the following. $\hat k_i$ is the unit vector for the $i$-direction.

The ultrasound attenuation coefficient $\alpha_{ij}(T)$ for the sound wave with the
polarization parallel to the $i$-direction and propagating along the $j$-direction is given by 
\begin{eqnarray}
{\alpha_{ij}(T)\over \alpha_{N}(T_{c1})}
&=&
{1\over 2T}\int^{\infty}_0d\omega  {\rm sech^2}{\omega\over 2T}A_{ij}(\omega,T)\label{eq:alpha}
\\
A_{ij}(\omega,T)
&=&
{1\over {\langle\Pi^2_{ij}\rangle}}
\langle\Pi^2_{ij}{1\over \omega}{\rm Re}\sqrt{\omega^2-|\Delta({\bf k},T)|^2}\rangle \label{eq:A}
\\
\Pi_{ij}
&=&
{\hat k_i}{\hat k_j}-{1\over 3}\delta_{ij}\label{eq:pi}
\end{eqnarray}
where $\alpha_N(T_{c1})$ is that in  the normal state at $T_{c1}$.
These formulae are standard ones except that (1) when treating non-unitary state,
we calculate these quantities for the up and down spin pairs independently,
and then sum over the two contributions.
This is permissible because these physical quantities heat and sound conserve the spin.
(2) According to Norman and Hirschfeld\cite{norman} the effects of inelastic scattering 
in the normal state are taken into account by scaling the results of 
Eqs.\ (\ref{eq:kappa}), and (\ref{eq:alpha}) by a factor$(1+bT_{c1}^2)/(1+bT^2)$
with $b$ being an adjustable parameter\cite{inela}.

%---------------------------------------------- fig.1--------
\begin{figure}
%%%\figureheight{1cm}
\mbox{\psfig{figure=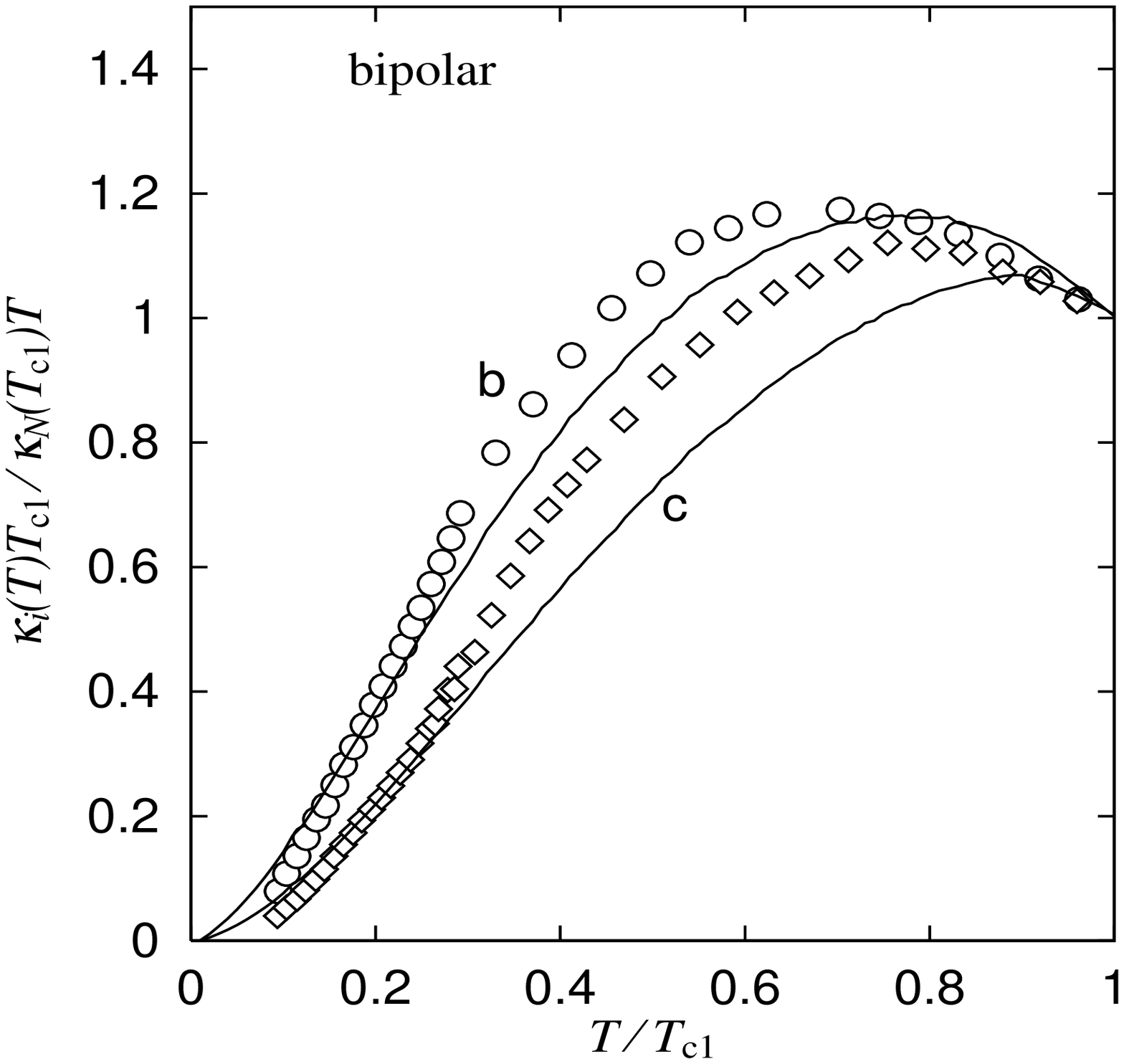,width=80mm}}\\ (a) \\
\mbox{\psfig{figure=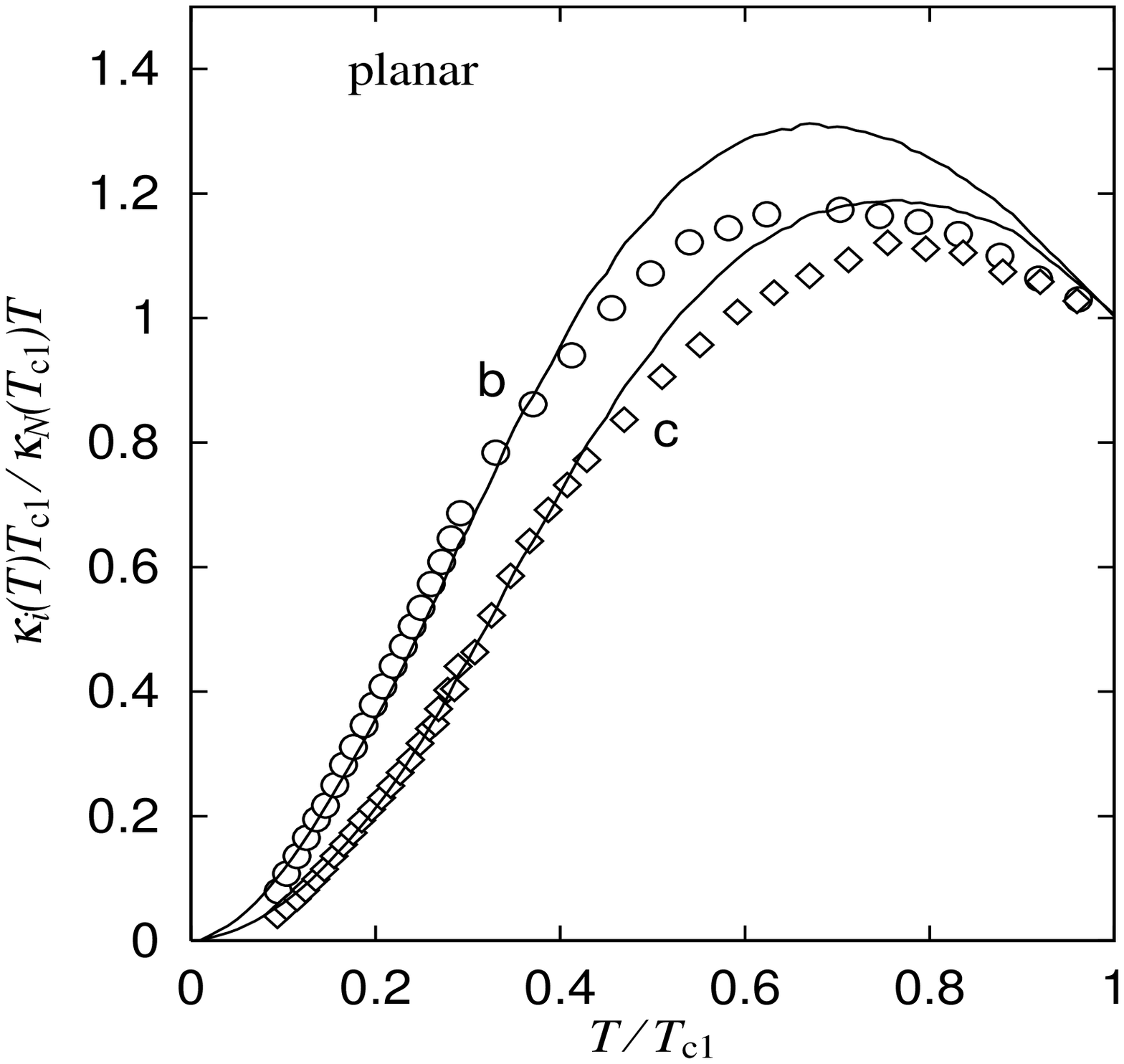,width=80mm}}\\ (b)
\caption{
The $T$-dependence of the thermalconductivities $\kappa_i(T)/T$
normalized by its value at $T=T_{c1}$ for the $b$-axis and $c$-axis.
(a) The bipolar state ($a_2=0.2$, $a_4=3.8$, $b_2=0.4$, $b_4=0$, and $b=5.0/K^2$), and 
(b) the planar state  ($a_2=4.4$, $a_4=0$, $b_2=1.8$, $b_4=0$, and $b=11.0/K^2$).
 The data are taken from Lussier at al\cite{lussier}.
}
\label{fig:thermal}
\end{figure}
%---------------------------------------------- fig.1--------

%---------------------------------------------- fig.2--------
\begin{figure}
%%%\figureheight{1cm}
%\vskip -5mm
\mbox{\psfig{figure=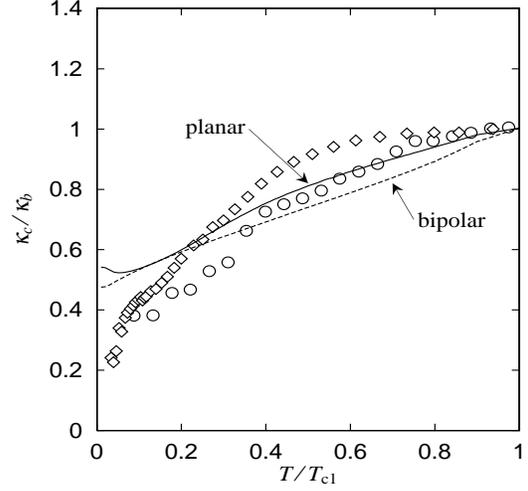,width=80mm}}
%\vskip 5mm
\caption{
The $T$-dependence of the anisotropy $\kappa_c/\kappa_b$
normalized at $T=T_{c1}$ for the bipolar (dashed line) and the planar
(solid line). The parameter values are the same as in Fig. \ref{fig:thermal}.
The data are taken from Lussier at al\cite{lussier} (circles) and
Suderow et al \cite{suderow} (diamonds).
}
\label{fig:thermalratio}
\end{figure}
%---------------------------------------------- fig.2--------

We show in Fig. \ref{fig:thermal} an example of the calculated results for the bipolar 
state (a) and the planar state (b) together with the experimental data by Lussier et al\cite{lussier}.
Although we do not exhaust all combinations of various parameters ($a_2,a_4,b_2,b_4,b$),
both fittings are equally good. 
In Fig. \ref{fig:thermalratio} the $T$-dependence of the thermalconductivity
anisotropy $\kappa_c/\kappa_b$ for both states are compared with the experimental data
by Lussier et al\cite{lussier} and Suderow et al\cite{suderow}.
It is rather difficult to decide which is better.

\begin{figure}
%% \figureheight{1cm}
\mbox{\psfig{figure=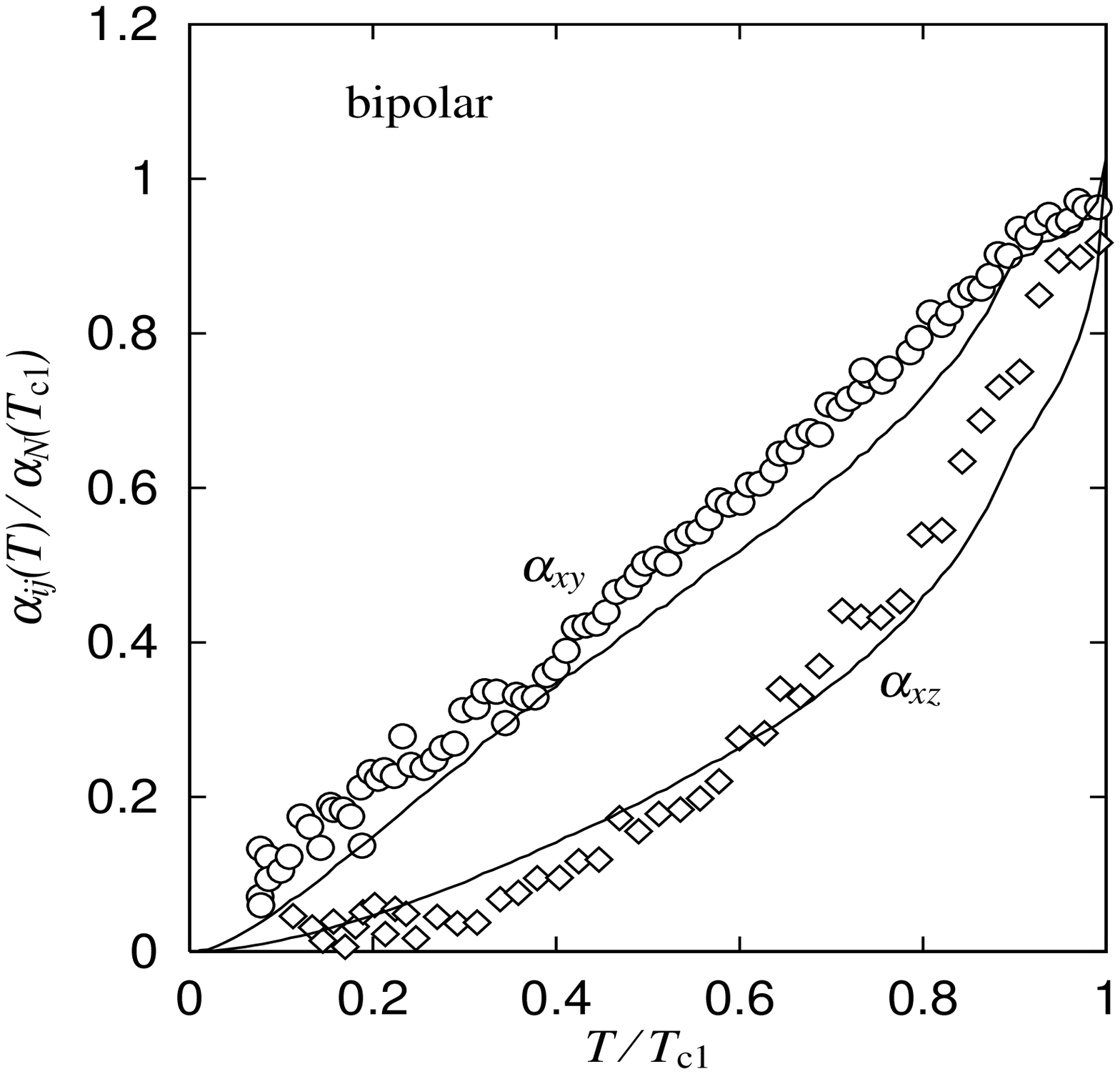,width=80mm}} \\ (a) \\
\mbox{\psfig{figure=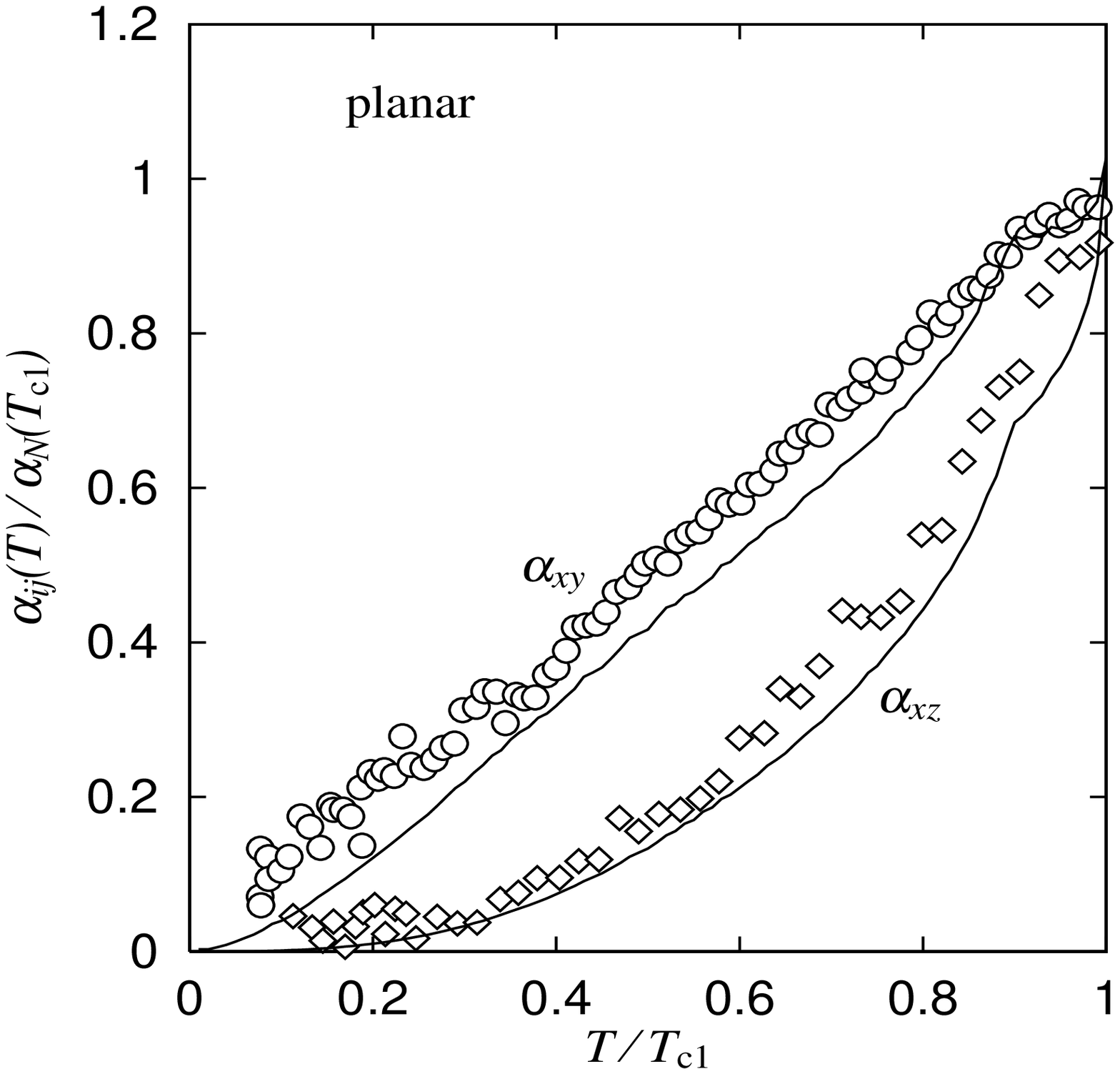,width=80mm}} \\ (b) 
\caption{
The $T$-dependence of the ultrasound attenuation coefficients
$\alpha_{xy}(T)$ and $\alpha_{xz}(T)$ normalized at $T=T_{c1}$.
(a) The bipolar state, and (b) the planar state.
The parameter values are the same as in Fig. 1.
The data are taken from Shivaram at al\cite{shivaram}.
}
\label{fig:ultra}
\end{figure}

The transverse sound attenuation coefficients for the two kinds
of the sound propagations are plotted in Fig. \ref{fig:ultra} (the bipolar state (a) and
the planar state (b)) with the same set of the parameter values as in the thermalconductivity and compared with the data by Shivaram et al\cite{shivaram}.
It is seen from Fig. \ref{fig:ultra}
that both theoretical curves for the bipolar and planar states  can explain the overall $T$-behaviors.
Also notice that the theoretical curves show a knee structure at the second transition $T_{c2}/T_{c1}=0.9$. This anomaly is clearly observed
by Taillefer et al\cite{taillefer} and Ellman et al\cite{ellman}, and not so clear in Shivaram et al's data cited here\cite{shivaram}.
We emphasize that although we did not hunt all the parameter space, 
both physical quantities $\kappa$ and $\alpha$ for two directions are consistently explained by the same set of the parameter values.

%---------------------------------------------- fig.3--------
%---------------------------------------------- fig.3--------

\section{Phase transitions}

Combining the orbital symmetry and the previous spin symmetry, we have completely determined 
the pairing symmetry realized in UPt$_3$.
We are now in position to explain the phase diagram observed experimentally although the essentially same explanation is given before\cite{our3}. The three basic 
assumptions are already adopted in this paper:
(1) The effective spin-orbit coupling felt by a Cooper pair is weak.
(2) The spin of the pairing is triple-degenerate.
(3) The antiferromagnetic fluctuations characterized by the triple-$Q$ vectors with
${\bf Q}_1=({1\over  2}, 0,0)$
and its equivalent positions ${\bf Q}_2$ and ${\bf Q}_3$ in reciprocal space 
($a^{\ast}=b^{\ast}={4\pi\over a\sqrt3}$, $c^{\ast}={2\pi\over c}$) in 
the hexagonal plane\cite{lussier} are responsible for breaking the triple degeneracy of the spin.

\begin{figure}
\begin{center}  \leavevmode
\mbox{\psfig{figure=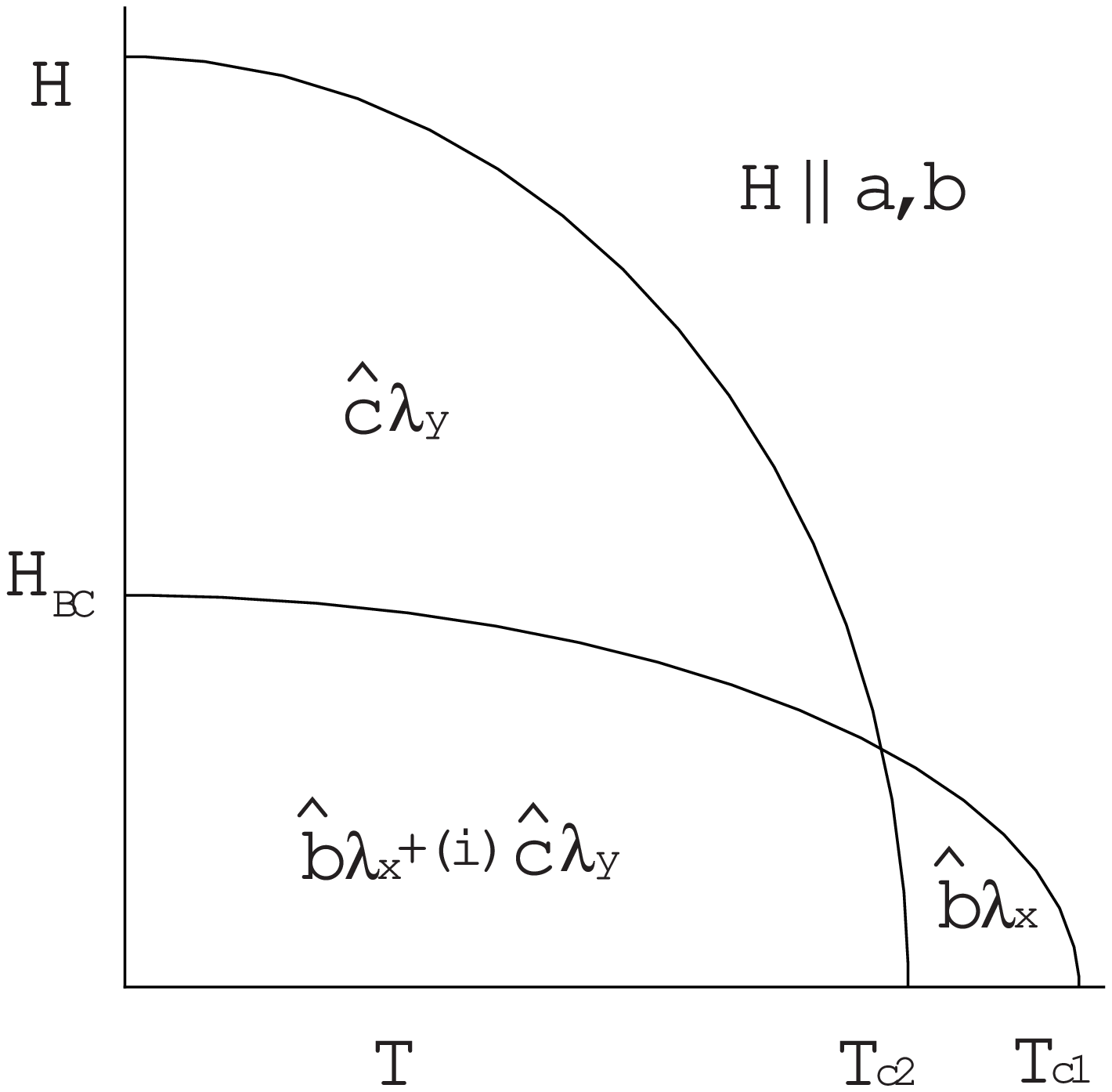,width=50mm}}\\
\mbox{\psfig{figure=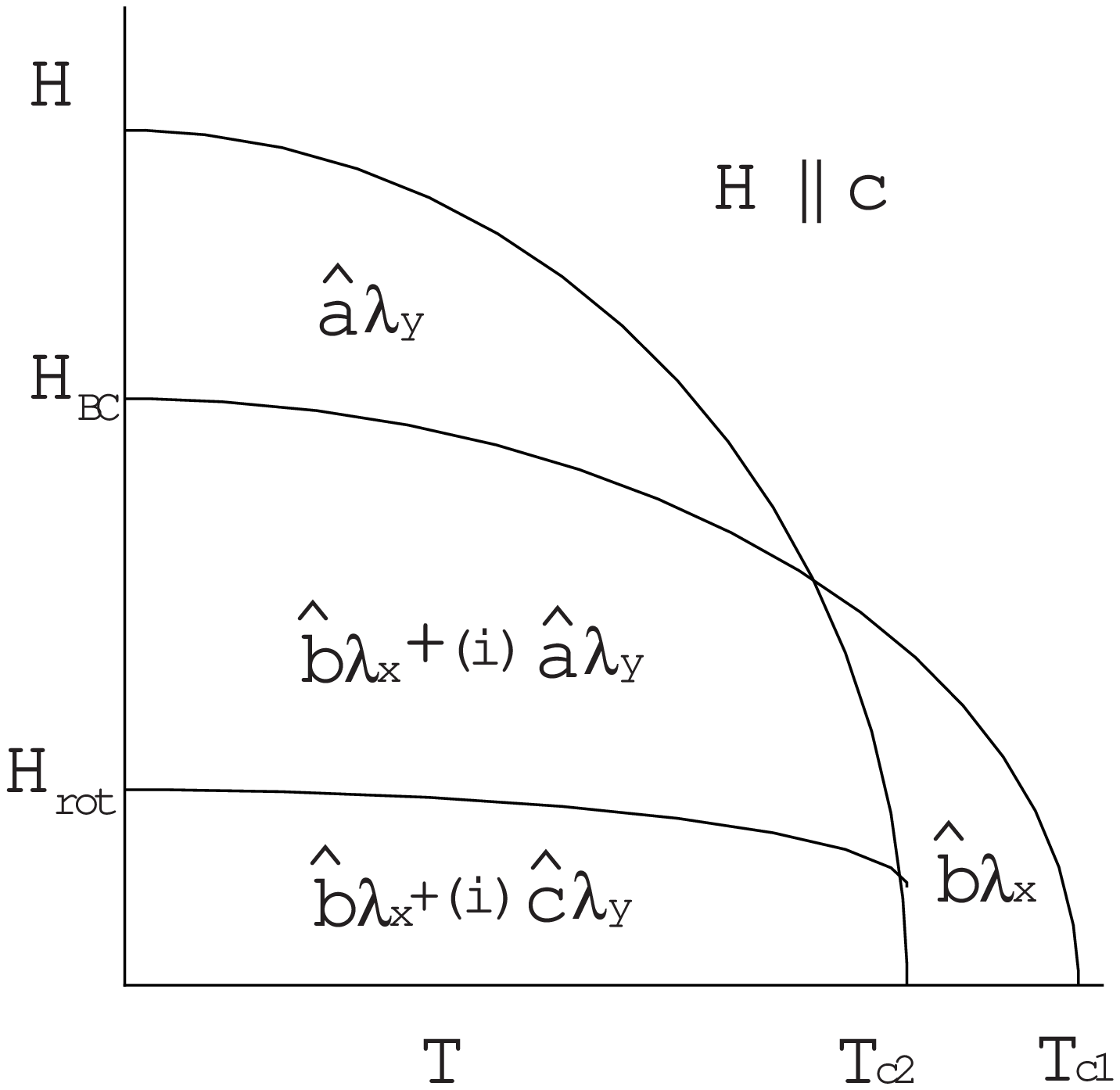,width=50mm}}
\end{center}
\caption{Schematic phase diagrams in $H$ vs $T$ for $H\parallel a, b$
 and $H\parallel c$ where the direction and the components of the $d$ vector
  for each phase are indicated. $H_{BC}$ is the second phase transition
   between the B and C phases.
$H_{rot}$ is the rotation field of the $\bf d$ vector
}
\label{fig:phasedia}
\end{figure}

These assumptions  lead to the following GL free energy
\begin{eqnarray}
F &=& F^{(2)}+F^{(4)}
\label{eq:gl0}
\\
F^{(2)} &=& \sum_{j=a,b,c}\alpha_0(T-T_{c}^j)|d_j|^2
\label{eq:gl2}
\end{eqnarray}
where $F^{(4)}$ is given in Appendix for the bipolar state; Eq. (\ref{eq:bi4})
and planar state; Eq. (\ref{eq:pla4}).
These particular quartic terms ensure that the bipolar or planar states is
stable in the ground state identified as the B phase and that the two
component spin vectors are orthogonal to each other.
Here we have taken the order parameter as $d_b={\hat b}\lambda_x({\bf k})$,
$d_c={\hat c}\lambda_y({\bf k})$ and 
$d_a={\hat a}\lambda_y({\bf k})$ for the planar state and as 
$d_b={\hat b}\lambda_x({\bf k})$,
$d_c=i{\hat c}\lambda_y({\bf k})$ and 
$d_a=i{\hat a}\lambda_y({\bf k})$ for the bipolar state.
The three transition temperatures $T^a_c>T^b_c>T^c_c$ which are originally degenerate are split into the two groups: 
$T^b_c$ and \{$T^a_c$ and $T^c_c$\} by the symmetry breaking field as mentioned. The latter two are further 
assumed to be slightly 
different: $T^a_c<T^c_c$ due to the small  uniaxial anisotropy of the system.
At $T=T^b_c$, which is identified as the upper critical temperature  $T_{c1}=0.58K$ the A phase characterized by $d_a=0, 
d_b\neq 0$ and $d_c=0$ appears first (see Fig. \ref{fig:phasedia})\cite{abm}.
Then at a lower temperature, $T=T^c_c$ which is identified as $T_{c2}=0.53K$ 
the second order transition from the A phase to the B phase which is characterized by either 
the non-unitary bipolar state or the unitary 
planar state.
It can be proved within the above free energy 
that third transition at $T=T^a_c$, which is designed to situate
a few mK below $T=T^c_c$
never realized at zero field. 

In order to discuss the phase diagrams for external field ${\bf 
H}$
applied parallel to and perpendicular to the $c$-axis.
We must take into account the vortex structure\cite{fujita} based on the 
above free energy functional by adding the terms describing the 
spatially varied order parameter:
\begin{eqnarray}
F &=& F_{grad}+F_{bulk}
\label{eq:gl00}
\\
F_{grad} &=& \sum_{j=a,b,c}\{K_1^j(|D_xd_j|^2+|D_yd_j|^2) \nonumber \\
&& \qquad  +K_2^j|D_zd_j|^2\}
\label{eq:glgr}
\\
F_{bulk} &=& \sum_{j=a,b,c}\alpha_0(T-T_{c}^j)|{d_j}|^2  \nonumber \\
&& \qquad +{1\over 2}\Delta\chi_P |{\bf H}\cdot {\bf d}|^2 +F^{(4)}.
\label{eq:glbulk}
\end{eqnarray}
The gradient term $F_{grad}$  describes the spatial variation of the order parameter  under external field 
with the gauge invariant derivative $D_j=-i\hbar\partial_j-{2e \over c}A_j$ ($\bf A$  is the vector potential).
Here there exist two kinds ($K_1^j$ and $K_2^j$ ($j=a,b$ and $c$)) of the gradient term associated with hexagonal symmetry where 
$K_1^b$ ($K_2^b$) can differ from $K_1^a$ and $K_1^c$ ($K_2^a$ and $K_2^c$) because of the symmetry breaking AF fluctuation,
that is, $K_1^b=K_1-\zeta {\bf M}^2, K_2^b=K_2-\zeta' {\bf M}^2$,
$K_1^a=K_1+\zeta {\bf M}^2,  K_2^a=K_2+\zeta' {\bf M}^2$,
$K_1^c=K'_1+\zeta {\bf M}^2,  K_2^c=K'_2+\zeta' {\bf M}^2$.
Here ${\bf M}^2$ is the amplitude 
of the AF fluctuation polarized along the $b$-axis.
It should be noted that there is no so-called 
gradient mixing term, which washes out 
the desired tetra-critical point,  in the present spin scenario.
This is quite 
different from that in the orbital scenarios\cite{sauls1,sauls2,joynt}
where the gradient mixing is inevitable. 

%---------------------------------------------- fig.4--------
%---------------------------------------------- fig.4--------

%---------------------------------------------- fig.5--------
\begin{figure}
\begin{center} \leavevmode
\mbox{\psfig{figure=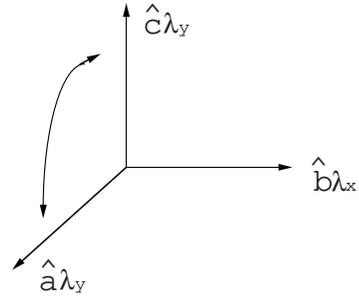,width=50mm}}
\end{center}
\caption{Schematic figure of the spin rotation under a weak $H$ field.
}
\label{fig:rotation}
\end{figure}
%---------------------------------------------- fig.5--------

The resulting phases in the $H$ vs $T$ shown in Fig. \ref{fig:phasedia} are 
fully characterized by specifying the $d$ vector in each phase, that is,
its direction, the number of the components
and the associated orbital parts.
The three vectors in the spin-triplet state are now
denoted by the real vectors $\hat{\bf a}$, $\hat{\bf b}$ and 
$\hat{\bf c}$ referring to the Cartesian coordinate in the hexagonal crystal. The high $T$ and low  (high)
field A(C) phase is described by a single component $d$ vector, while the B phase in the low $T$ and low $H$ is a non-unitary bipolar state or unitary planar state characterized by two-component $d$ vector. 
Because of the $\Delta\chi_P$ term in the above free energy Eq.(\ref{eq:glbulk}), which tends to align the $d$ vector perpendicular to ${\bf 
H}$, the B phase for ${\bf H}\parallel {\bf c}$ is further subdivided into 
$\hat{\bf b}\lambda_x({\bf k})+(i)\hat{\bf c}\lambda_y({\bf k})$ in low $H$ and $\hat{\bf b}\lambda_x({\bf k})+(i)\hat{\bf a}\lambda_y({\bf k})$ in high $H$. As $H(\parallel {\bf c})$ 
increases the $d$ vector rotates,
corresponding to the Knight shift behavior\cite{tou} at $H\sim$2kG.
The $d$ vector rotation is schematically shown in Fig.\ref{fig:rotation} where
one of the $d$ vector components rotates upon increasing $H(\parallel c)$ while keeping
the other component fixed to the $b$-axis.

\section{Summary and conclusions}
Before going to a summary and conclusions,  we give 
additional discussions to clear our standpoints and to answer possible objections
raised in several places.\cite{heffner,sauls2,sauls3}

\subsection{Supplemental discussions}

\underline {The double transition}: The origin of the $T_c$-splitting comes from the crystalline hexagonal symmetry breaking due to either incommensurte lattice modulations, or stacking faults along the $c$-axis occurring from a high temperature, which are well correlated with the existence of the $T_c$-splitting\cite{midgley}.
This leads to the different spectral weights of the otherwise triply degenerate antiferromagnetic fluctuations whose polarization points to the $a^{\ast}(\parallel b)$-axis, thus the originally three equivalent $a^{\ast}$-axes become inequivalent from  a high $T$. This inequivalence is indeed observed in Tou et al's Knight shift experiment\cite{touprivate}.
The spectral weight difference of the antiferromagnetic fluctuations breaks the degeneracy of the multiple component order parameters. 
The growing evidence\cite{nmr,visser,schottl,yaouanc} shows that the antiferromagnetism is not true static and long-range order, which is unable to work as a symmetry-breaker as postulated by the E$_{1g}$ and E$_{2u}$
scenarios. This is a main reason why we abandoned our original 
scenario\cite{original}. Moreover, the expected antiferromagnetic
moment rotation upon applying the basal plane magnetic field is not
observed by the neutron experiment\cite{neutron}. This makes these 
scenarios more difficult because the basal plane phase diagram 
is anisotropic which is not seen experimentally.
As for the main pairing mechanism we think the observed ferromagnetic 
fluctuations\cite{bern} for stabilizing the triplet channel as in superfluid $^3$He while the antiferromagnetic fluctuations mentioned above are merely to split the transition temperature with otherwise degenerate state. Indeed the latter fluctuation intensity is
known to decrease when the system undergoes superconducting\cite{aeppli}, revealing it workable as a
symmetry breaker.

\underline{The spin rotation}: It was remarkable to observe that the spin susceptibility begins to decrease as $H(\parallel c)$ increases, implying that the $d$-vector
rotates at around 2kG  within the B phase 
upon responding to external field to lower the Zeeman energy.
This immediately leads us that 
(1)the effective spin-orbit coupling felt by a Cooper pair is weak
contrary to what has been claimed\cite{anderson}.
(2)The spin component of the order parameter must be multiple,
negating not only the spin-singlet scenario E$_{1g}$\cite{joynt}, but also the single 
spin component scenario such as E$_{2u}$, which also assumes the strong spin-orbit coupling\cite{sauls1,sauls2,sauls3}.
Our scenario is able to explain this remarkable observation as explained before.

\underline{The $H_{c2}$ crossover phenomenon}:  The upper critical field $H^{\perp}_{c2}$ for the basal plane exceeds that for the $c$-axis at low $T$. The $T$-dependence of $H^{\perp}_{c2}$ exhibits no
saturation towards $T\rightarrow0$. This crossover behavior of $H^{\perp}_{c2}$ and $H^{\parallel}_{c2}$ is often taken as evidence\cite{choi} of the absence of the Pauli limiting of $H^{\perp}_{c2}$. We think, instead,
that the anomalous $H^{\perp}_{c2}$ curve indicates a new phase  above the C phase. The recent Knight shift experiment by Tou et al\cite{touprivate} shows that $\chi(H)$  begins to decrease as $H(\parallel a)$ increases at around 10kG in the middle of the C phase.

\subsection{Conclusions}

We have demonstrated that the proposed orbital part combined by the previously determined spin part, namely the non-unitary 
bipolar state and unitary planar state gives rise to a reasonable explanation 
for the overall $T$-dependences of the thermalconductivity and transverse ultrasound attenuation both for the $c$-axis and the basal plane.
These states can also explain the major previously known thermodynamic and transport experiments, including
the specific heat ($\sim T^2$), the nuclear relaxation rate ($\sim T^3$)
and the longitudinal ultrasound attenuation for both directions ($\sim T$), all of which merely point to the existence of a line node since the former two are directionally averaged quantities. The direction-resolved 
anisotropic penetration depth probed by $\mu$SR\cite{penetration}
for two direction
($\parallel c$ and $\perp c$) is interesting to interpret in terms of these states, which belongs to a future study. Other measurements related to surface sensitive technique such as point contact tunneling 
spectroscopy\cite{goll}  and Josephson tunneling\cite{sumiyama} which yield valuable information of each phase A, B and C phases
remain to be done although we believe that our bipolar or planar states 
are basically able to understand these data.

It is rather difficult to distinguish between the bipolar and planar states both theoretically and experimentally. The existing Knight shift experiments are consistent completely with both states. There is no distinction between them. As for the orbital part related to the gap 
topology in the B phase the bipolar state has four line nodes parallel to the $c$-axis while the planar state has two quadratic point nodes at the poles in addition to a line node on the equator in common. Thus we expect that there exists the in-plane four-fold anisotropy for the bipolar state while there is no in-plane anisotropy 
for the planar state. We warn, however, that this  anisotropy may not be so strong to be able to discriminate them because of the background hexagonal symmetry. But this kind of experiments,
including the transverse sound attenuation for various in-plane polarizations under the sound propagations kept along the $c$-axis are worth trying. In connection with the $H_{c2}$ crossover 
phenomenon mentioned above the new phase transition which appears when $H(\parallel a)\sim$ 10kG should be detected by other phase transition sensitive experiments such as elastic anomaly.\cite{sawada} It is also interesting to further investigate the vortex structure \cite{knigavko} in terms of the 
identified state.

\appendix
\section{Stability conditions of the bipolar and planar states}

We start with the general free energy functional valid up to the quartic terms\cite{jones} 
invariant under spatial and spin rotations and a guage 
transformation given by 
\begin{eqnarray}
f &=& f^{(2)}+f^{(4)}
\label{eq:ftotal}
\\
f^{(2)} &=& \alpha_0(T-T_{c}){\rm tr}AA^{\dagger}
\label{eq:f2}
\\
f^{(4)} &=&
\beta_1 {\rm tr}|A{\tilde A}|^2
+\beta_2 ({\rm tr}|AA^{\dagger}|)^2                  \nonumber \\ &&
+\beta_3 {\rm tr}[(A^{\dagger}A)(A^{\dagger}A)^{\ast}]       \nonumber  \\ &&
+\beta_4 {\rm tr}(AA^{\dagger})^2
+\beta_5 {\rm tr}[(AA^{\dagger})(AA^{\dagger})^{\ast}]
\label{eq:f4}
\end{eqnarray}
where $A_{\mu i}$ is a complex 3$\times$3 matrix. $\mu (=1,2,3)$ and $i(=x,y,z)$
denote the spin and orbital indices respectively.
${\tilde A}$ and $A^{\dagger}$ are the transpose and Hermitian conjugate of $A$.
In the case of the $p$-wave pairing such as superfluid He$^3$, $A$
relates to the $d$-vector as ${\bf d}({\bf k})=A\cdot{\hat {\bf k}}$.
When the orbital component is confined to just two, whose basis functions are $\lambda_x({\bf k})$
and $\lambda_y({\bf k})$, 
we can reduce the $d$-vector  as 
${\bf d}({\bf k})={\vec A}_x\lambda_x({\bf k})+{\vec A}_y\lambda_y({\bf k})$.

By writing the $d$-vector as ${\vec A}_x={\vec \alpha}+i{\vec \beta}$, and 
${\vec A}_y={\vec \gamma}+i{\vec \delta}$
where ${\vec \alpha}$, ${\vec \beta}$, ${\vec \gamma}$ and ${\vec \delta}$ are real vectors,
we find from Eq.\ (\ref{eq:f4})
\begin{eqnarray}
f^{(4)}
&=&
\beta_1\{({\vec \alpha}\cdot{\vec \alpha}
-{\vec \beta}\cdot{\vec \beta}
+{\vec \gamma}\cdot{\vec \gamma}
-{\vec \delta}\cdot{\vec \delta})^2          \nonumber \\ && \quad
+4({\vec \alpha}\cdot{\vec \beta}+{\vec \gamma}\cdot{\vec \delta})\} 
\nonumber \\&&
+\beta_2 (
{\vec \alpha}\cdot{\vec \alpha}
+{\vec \beta}\cdot{\vec \beta}
+{\vec \gamma}\cdot{\vec \gamma}
+{\vec \delta}\cdot{\vec \delta}
)^2
\nonumber \\ &&
+\beta_3 \{({\vec \alpha}\cdot{\vec \alpha}+{\vec \beta}\cdot{\vec \beta})^2
+({\vec \gamma}\cdot{\vec \gamma}+{\vec \delta}\cdot{\vec \delta})^2         \nonumber \\ && \quad
+2({\vec \alpha}\cdot{\vec \gamma}+{\vec \beta}\cdot{\vec \delta})^2
-2({\vec \beta}\cdot{\vec \gamma}-{\vec \alpha}\cdot{\vec \delta})^2\}
\nonumber \\ &&
+\beta_4 \{({\vec \alpha}\cdot{\vec \alpha}+{\vec \beta}\cdot{\vec \beta})^2
+({\vec \gamma}\cdot{\vec \gamma}+{\vec \delta}\cdot{\vec \delta})^2         \nonumber \\ && \quad
+2({\vec \alpha}\cdot{\vec \gamma}+{\vec \beta}\cdot{\vec \delta})^2
+2({\vec \beta}\cdot{\vec \gamma}-{\vec \alpha}\cdot{\vec \delta})^2\}
\nonumber \\ &&
+\beta_5 \{({\vec \alpha}\cdot{\vec \alpha}-{\vec \beta}\cdot{\vec \beta})^2
+({\vec \gamma}\cdot{\vec \gamma}-{\vec \delta}\cdot{\vec \delta})^2         \nonumber \\ && \quad
+4({\vec \alpha}\cdot{\vec \beta})^2+4({\vec \gamma}\cdot{\vec \delta})^2
+2({\vec \alpha}\cdot{\vec \gamma}-{\vec \beta}\cdot{\vec \delta})^2         \nonumber \\ && \quad
+2({\vec \beta}\cdot{\vec \gamma}+{\vec \alpha}\cdot{\vec \delta})^2\}.
\label{eq:f}
\end{eqnarray}
In order to know the stability of the bipolar and planar states, let us consider small deviations from the respective states: 
$A_{\mu i}=A_{\mu i}^0+a(\mu,i)+ib(\mu,i)$ with $i=x,y$.
For the bipolar state $f_{bipolar}$ is written  up to the quadratic order in small deviations as 
\begin{eqnarray}
\delta f_{bipolar}
&=&
\beta_1\{8a^2(1,x)+2(a(2,y)+b(1,x))^2\}
\nonumber
\\ &&
+\beta_3\{4a^2(1,x)+(a(1,y)+b(2,x))^2     \nonumber \\ && 
\quad -(a(2,x)+b(1,y))^2\}
\nonumber
\\ &&
+\beta_4\{4a^2(1,x)+(a(1,y)+b(2,x))^2     \nonumber \\ && 
\quad +(a(2,x)+b(1,y))^2\}
\nonumber
\\ &&
+\beta_5\{4a^2(1,x)-(a(1,y)+b(2,x))^2     \nonumber \\ && 
\quad +(a(2,x)+b(1,y))^2     \nonumber \\ && 
\quad -2a^2(3,y)-2b^2(3,x)\}.
\label{eq:deltaf}
\end{eqnarray}
This leads to the stability conditions for the bipolar state:
\begin{eqnarray}
&& \beta_1>0, \ \ \ \beta_5<0, \ \ \  2\beta_1+\beta_3+\beta_4+\beta_5>0
\nonumber \\
&& \beta_3+\beta_4-\beta_5>0, \ \ \ -\beta_3+\beta_4+\beta_5>0.
\label{eq:sta1}
\end{eqnarray}
For the planar state  $\delta f_{planar}$ is calculated as 
\begin{eqnarray}
\delta f_{planar}
&=&
\beta_1\{-2(b(1,x)-b(2,y))^2-4b^2(1,y)      \nonumber \\ &&
\quad   -4b^2(2,x)-4b^2(3,x)-4b^2(3,y)\}
\nonumber
\\ && 
+\beta_3\{4a^2(1,x)+(a(1,y)+b(2,x))^2         \nonumber \\ &&
\quad   -(b(1,y)-b(2,x))^2\}
\nonumber
\\ && 
+\beta_4\{4a^2(1,x)+(a(1,y)+a(2,x))^2         \nonumber \\ &&
\quad   +(b(1,y)-b(2,x))^2\}
\nonumber
\\ && 
+\beta_5\{4a^2(1,x)+(a(1,y)+a(2,x))^2         \nonumber \\ &&
\quad   -(b(1,y)-b(2,x))^2                 \nonumber \\ &&
\quad   -2b^2(3,x)-2b^2(3,y)\}.
\label{eq:deltaf2}
\end{eqnarray}
The stability conditions for the planar state is given by
\begin{eqnarray}
&& \beta_1<0, \  \   \ \beta_5<0,  \nonumber \\
&& \beta_3+\beta_4+\beta_5>0
\nonumber
\\ 
&& -\beta_3+\beta_4-\beta_5>0,
\label{eq:sta2}
\end{eqnarray}

The resulting quartic terms for each state is rewritten as 
\begin{eqnarray}
f^{(4)}_{bipolar}
&=&
(\beta_1+{1\over2}\beta_{345})
({\vec \alpha}\cdot{\vec \alpha}-{\vec \delta}\cdot{\vec \delta})^2         \nonumber 
\\ &&
+(\beta_2+{1\over2}\beta_{345})({\vec \alpha}\cdot{\vec \alpha}+{\vec \delta}\cdot{\vec \delta})^2
\nonumber \\ &&
+2(-\beta_3+\beta_{45})({\vec \alpha}\cdot{\vec \delta})^2
\label{eq:bi4}
\\
f^{(4)}_{planar}
&=&
(\beta_{12}+{1\over2}\beta_{345})({\vec \alpha}\cdot{\vec \alpha}+{\vec \gamma}\cdot{\vec \gamma})^2
\nonumber \\ && \!\! 
+{1\over 2}\beta_{345}\{({\vec \alpha}\cdot{\vec \alpha}-{\vec \gamma}\cdot{\vec \gamma})^2
+4({\vec \alpha}\cdot{\vec \gamma})^2\}
\label{eq:pla4}
\end{eqnarray}
where $\beta_{ijk\cdots}=\beta_i+\beta_j+\beta_k+\cdots$.
The minimization is found to require from Eqs.\ (\ref{eq:bi4}) and (\ref{eq:pla4})
that ${\vec \alpha}\perp{\vec \delta}$ and $|{\vec \alpha}|=|{\vec \delta}|$ for the bipolar state
and ${\vec \alpha}\perp{\vec \gamma}$ and $|{\vec \alpha}|=|{\vec \gamma}|$ for the planar state.
Namely, the spin component vectors for each orbital component are orthogonal to each other.
The general structures of the 3$\times$3 matrix $A$ under a fixed $\alpha$ vector parallel 
to the $1$-direction are now written as 
\begin{eqnarray}
A_{bipolar}= \left(
                                      \begin{array}{ccc}
                                       \alpha_1 & 0    & 0 \\
                                         0 & i\delta_2 & 0 \\
                                         0 & i\delta_3 &0 \\
                                    \end{array}
                                     \right) \label{eq:abipolar}
\end{eqnarray}
\begin{eqnarray}
A_{planar}= \left(
                                      \begin{array}{ccc}
                                       \alpha_1 & 0    & 0 \\
                                         0 & \gamma_2 & 0 \\
                                         0 & \gamma_3 &0 \\
                                    \end{array}
                                     \right) \label{eq:aplanar}
\end{eqnarray}
where $\alpha_1$, $\delta_2$, $\delta_3$, $\gamma_2$ and $\gamma_3$ are real numbers.
In the main text the spin indices $\mu$=1, 2, 3 correspond to the $b$, $c$ and $a$-axes respectively.


\begin{thebibliography}{99}
\bibitem{E1u}
The E$_{1u}$ state is given by $\lambda_x(k)=k_x$ and $\lambda_y(k)=k_y$ which has no
line node on the equator in general both for the bipolar or planar states. Thus it is not appropriate for the present case.
\bibitem{same}The gap topology of the planar state is the same as in 
Sauls state. We note, however, that the spin structure is completely different.
\bibitem{abm}In addition to the stability conditions given by
the Appendix for the planar and bipolar states as the B phase
we need an additional condition that the A phase is correctly described by a polar type state: $d_b={\hat b}\lambda_x({\bf k})$, not
an ABM type state: $d_b={\hat b}(\lambda_x({\bf k})+i\lambda_y({\bf k}))$. Since the quartic terms of these states are written as $f^{(4)}_{polar}=\beta_{12345}$ and $f^{(4)}_{ABM}=\beta_{245}$, $\beta_1+\beta_3<0$
is required.
\end{thebibliography}
\end{document}